\documentclass[12pt]{article}
\usepackage{graphicx}

\begin{document}

\title{Numerical analyses of emission of a single-photon pulse based on single-atom cavity quantum electrodynamics}

\author{Hiroo Azuma\thanks{Present address: Nisshin-scientia Co., Ltd., 8F Omori Belport B, 6-26-2 MinamiOhi, Shinagawa-ku, Tokyo 140-0013, Japan.
Email: hiroo.azuma@m3.dion.ne.jp}
\\
\\
{\small Advanced Algorithm \& Systems Co., Ltd.,}\\
{\small 7F Ebisu-IS Building, 1-13-6 Ebisu, Shibuya-ku, Tokyo 150-0013, Japan}\\
}

\date{\today}

\maketitle

\begin{abstract}
We numerically investigate an on-demand single-photon source,
which is implemented with a strongly coupled atom-cavity system, proposed by Kuhn {\it et al}., Appl. Phys. B \textbf{69}, 373 (1999).
In the scheme of Kuhn {\it et al}.,
a $\Lambda$-type three-level atom is captured in a single-mode optical cavity.
Considering the three atomic levels,
the ground state $u$, the first excited state $g$ accompanying the cavity mode,
and the second excited state $e$, in the $\Lambda$-configuration,
we assume that a classical field and a quantized cavity field lead to the transition between $u$ and $e$ and that between $e$ and $g$, respectively.
The classical light pulse rising sufficiently slowly triggers an adiabatic process of the system and lets a single photon of the cavity mode emerge.
We simulate this adiabatic evolution and transmission of the single photon through an imperfect mirror of the cavity using the master equation.
We concentrate on examining physical properties of the efficiency of single-photon generation,
the fluctuation of the duration of the photon emission,
and the time of the emission measured from a peak of the trigger pulse.
We find a function that approximates to the efficiency closely and the upper bound of the fluctuation of the duration.
\end{abstract}

\section{\label{section-introduction}Introduction}
To perform quantum information processing,
such as quantum cryptography and quantum computation,
with photons,
we often have to prepare an on-demand single-photon source.
That is to say,
we have to generate a single photon at arbitrarily chosen time with high efficiency.
For example,
the Bennett-Brassard 84 (BB84) protocol requires polarized single photons as flying quantum bits (qubits)
\cite{Bennett1984}.
The Ekert 91 (E91) protocol also needs pairs of entangled single photons to generate nonlocal correlations
\cite{Ekert1991}.
Moreover,
Knill, Laflamme, and Milburn's conditional sign-flip gate works using single photons as dual-rail qubits
\cite{Knill2001}.
Thus, the single-photon emitter is regarded as one of the most important components for constructing quantum information processors.
However, so far, practical realization of the on-demand single-photon gun has not been established yet.
Since various protocols of quantum cryptography \cite{Bennett1984,Ekert1991}
and algorithms of quantum computation \cite{Shor1997,Grover1997} appeared,
many researchers have been trying to develop the deterministic single-photon source with a wide variety of physical systems.

If we attempt to fabricate a deterministic single-photon gun from weak laser light primitively,
the following two problems confront us.
The first one is fluctuation of the number of photons emitted together at the same time.
A coherent state $|\alpha\rangle$ provides $\langle\hat{n}\rangle=|\alpha|^{2}$ and $\Delta n^{2}=|\alpha|^{2}$
as the expectation value and the variance of the operator for the number of photons, respectively.
Thus, photon bunching hinders the realization of the single-photon emission.
The second one is fluctuation of the time interval for photon emission that obeys the Poisson distribution.
This stochastic property prevents us from building the on-demand source.
Hence, construction of the on-demand single-photon source is a difficult and challenging problem.

Implementation of the single-photon source with a quantum dot is discussed both theoretically and experimentally
\cite{Benson2000,Michler2000,Santori2001,Pelton2002,Pelton2003,Kiraz2004}.
Recently, the solid-state single-photon source has attracted researchers' attention.
In Ref.~\cite{Bernien2012},
emission of two photons from separate nitrogen vacancy centres in diamond was observed.
In Ref.~\cite{Rogers2014},
the single-photon source was built with a negative silicon vacancy centre in diamond.
Fabrication of the single-photon source from an atom-cavity system has been studied by some groups not only theoretically but also experimentally
\cite{Brattke2001,Mucke2013}.

Kuhn {\it et al}. have discussed the method for realizing the on-demand single-photon source that is built with a strongly coupled atom-cavity system
\cite{Kuhn1999}.
This method is an advanced version of the proposals of Law {\it et al}. \cite{Law1996,Law1997}.
In Ref.~\cite{Law1996},
Law and Eberly studied a $\Lambda$-type three-level atom interacting with a cavity mode and a classical driving field.
In Ref.~\cite{Law1997},
Law and Kimble proposed a scheme for generating a single-photon state transmitted out of an optical cavity,
in which the $\Lambda$-type three-level atom was captured.

In the above proposals of Law {\it et al}., we consider the three atomic levels,
the ground state $u$, the first excited state $g$ accompanying the cavity mode, and the second excited state $e$, in the $\Lambda$-configuration.
A classical field and the quantized cavity field provoke the transition between $u$ and $e$ and that between $e$ and $g$, respectively.
By injecting the classical light as a trigger pulse into the cavity,
we can excite an initial state at the level $u$ into the intermediate state at the level $e$.
After giving rise to the cavity mode from the transition between the intermediate state $e$ and the state at the level $g$,
we can let the single photon in the cavity mode pass through an imperfect mirror of the cavity.

Kuhn {\it et al}. have introduced an adiabatic process into the above model to avoid spontaneous emission of light
from the intermediate state and caused emergence of the cavity mode theoretically
by applying the classical trigger pulse to the system.
To assure the adiabatic process,
we must make the trigger pulse rise sufficiently slowly \cite{Kuhn1999}.
This process is called the stimulated Raman adiabatic passage (STIRAP) \cite{Vitanov2017}.
The method of Kuhn {\it et al}. has been demonstrated in laboratories according to
Refs.~\cite{Hennrich2000,Kuhn2002,Hennrich2004,Hijlkema2007}.
In particular,
Keller {\it et al}. have performed the scheme of Kuhn {\it et al}. under nearly ideal conditions using a calcium ion
\cite{Keller2003,Keller2004a,Keller2004b}.

In the present paper,
we numerically investigate the dynamics of the deterministic single-photon source proposed by Kuhn {\it et al}.
We analyse the adiabatic evolution of the atom-cavity system and transmission of a high single-photon flux through the cavity mirror
by using the master equation.
We focus on the following three topics.
The first one is the efficiency of single-photon generation.
It is proportional to both an expectation value of the number of photons in the cavity mode
and a decay rate that governs the transmission of the photon through the mirror.
With an empirical manner, we find a function that approximates to the efficiency closely.
The second one is the fluctuation of the duration of the photon emission.
We calculate full width at half maximum of the time evolution of the probability that the photon in the cavity mode emerges and regard it as the fluctuation.
We find its upper bound analytically by the adiabatic approximation.
The third one is the time of the emission of the photon measured from a peak of the trigger pulse.
As the decay rate increases,
the single photon is emitted earlier with respect to the peak of the trigger pulse.
We estimate the time of the emission numerically.

Here, we emphasize the prior works concerning our study closely.
In Ref.~\cite{Keller2004b},
single-photon pulses for different pump laser profiles,
for instance, Gaussian pumps and a square-wave pump,
were examined
and the suppression of two-photon events was confirmed.
In this work, Keller {\it et al}. demonstrated the good agreement between the theoretical model and the experimental data.
Figure 2a of Ref.~\cite{Keller2004b} showed the following.
The time when the waveform became maximum preceded the peak of the classical trigger pulse in the experimental data.
This fact is the third topic treated in the current paper and is discussed in Sect.~\ref{section-t-max}.
In Refs.~\cite{Vasilev2010,Khanbekyan2017},
proper shapes of the classical trigger pulse for emission of single-photon wave packets of a desired shape with high efficiency
were investigated.
By contrast, in Sect.~\ref{section-trigger-pulse} of the present paper, we assume that the shape of the input pulse is given by the Gaussian form.

This paper is organized as follows.
In Sect.~\ref{section-review-Kuhn}, we give a review of the method proposed by Kuhn {\it et al}. for emitting the single photon.
In Sect.~\ref{section-trigger-pulse}, we define the classical trigger pulse and derive an explicit form of the adiabaticity constraint.
In Sect.~\ref{section-master-equation}, we introduce the master equation that describes the single-photon emission.
In Sect.~\ref{section-numerical-calculation-populations}, we show time evolution of each state of the system
during the adiabatic process.
In Sect.~\ref{section-efficiency}, we examine the efficiency of the single-photon generation.
We find the function which approximates to the efficiency well.
In Sect.~\ref{section-time-fluctuation}, we investigate the fluctuation of the duration of the photon emission and find its upper bound.
In Sect.~\ref{section-t-max}, we show variation of the time when the photon is emitted.
Section~\ref{section-discussion} gives brief discussion.
In Appendix~\ref{section-numerical-calculation-master-equation},
we explain how to solve the master equation numerically.

\section{\label{section-review-Kuhn}A review of the method proposed by Kuhn {\it et al}. for emitting a single photon}
\begin{figure}
\begin{center}
\includegraphics{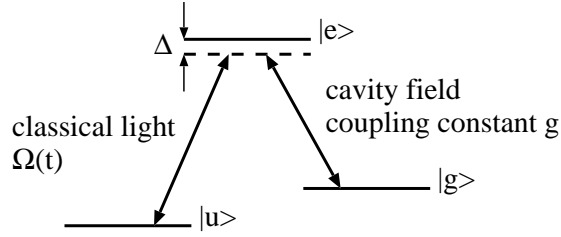}
\end{center}
\caption{An energy-level diagram for a $\Lambda$-type three-level atom,
whose ground and excited states are given by $|u\rangle$, $|e\rangle$, and $|g\rangle$.
The optical Bloch equations describe a transition between $|u\rangle$ and $|e\rangle$.
The Jaynes-Cummings interaction governs a transition between $|e\rangle$ and $|g\rangle$
with the cavity field.}
\label{Figure01}
\end{figure}

In this section,
we give a review of the scheme proposed by Kuhn {\it et al}. for generating an on-demand single photon \cite{Kuhn1999}.
We consider a $\Lambda$-type three-level atom,
whose ground and excited states are represented by $|u\rangle$, $|e\rangle$, and $|g\rangle$ as shown in Fig.~\ref{Figure01}.
First, we assume that the transition between $|u\rangle$ and $|e\rangle$ is induced by a classical light whose frequency and amplitude are given by
$[(E_{0}/\hbar)-\Delta]$ and $\Omega(t)$, respectively.
To describe the time evolution of the transition between $|u\rangle$ and $|e\rangle$,
we employ the optical Bloch equations.
Second, we assume that the transition between $|g\rangle$ and $|e\rangle$ is caused by the Jaynes-Cummings interaction
with the coupling constant $g$ and the cavity mode
whose frequency is given by $\omega$.
In both transitions,
we put $\Delta$ as the common detuning of the classical light and the cavity field from the intermediate level $|e\rangle$.

Here, we introduce a number state of photons in the cavity mode as $|n\rangle$ for $n=0, 1, 2, ...$.
Then, we consider three states,
$|u,0\rangle$, $|e,0\rangle$, and $|g,1\rangle$.
The states $|u,0\rangle$ and $|e,0\rangle$ are coupled by the classical light.
The states $|e,0\rangle$ and $|g,1\rangle$ are coupled by the cavity mode.
The essence of the scheme proposed by Kuhn {\it et al}. is an adiabatic process which lets the initial state $|u,0\rangle$ evolve into the state $|g,1\rangle$
without going through the intermediate state $|e,0\rangle$.
Thus, we can avoid spontaneous emission of the classical light that is due to the transition from the excited state $|e,0\rangle$ to the ground state $|u,0\rangle$.
If the system reaches the state $|g,1\rangle$,
the subsequent decay of the cavity mode causes the single-photon emission and the system settles itself in the state $|g,0\rangle$.
In order to let the atom-cavity system pursue the adiabatic process,
we have to apply the classical trigger pulse rising sufficiently slowly to the system.

As mentioned above, after the emission of the single photon,
the state of the system changes into $|g,0\rangle$.
Then, we apply a repumping pulse to the atom-cavity system and bring the system back to the initial state $|u,0\rangle$.
Repeating this cycle, we obtain a bit-stream of the single photons.

\begin{figure}
\begin{center}
\includegraphics{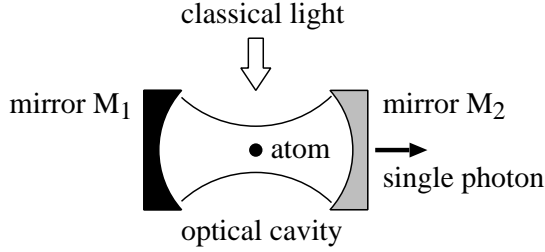}
\end{center}
\caption{A schematic diagram of the single-photon source.}
\label{Figure02}
\end{figure}

Because the decay of the cavity mode generates the emission of the single photon,
the efficiency of the emission is proportional to the decay rate.
At the same time, the efficiency has to be proportional to an expectation value of the number of photons in the cavity mode.
Figure~\ref{Figure02} illustrates the final form of the system that emits the single photon.
The left mirror $M_{1}$ is perfect and it reflects a single photon with a probability of unity.
By contrast, the right mirror $M_{2}$ is not perfect and the single photon passes through it with the decay rate.

Here, we consider an explicit form of the Hamiltonian that describes the above atom-cavity system.
We write down the states $|u\rangle$, $|e\rangle$, and $|g\rangle$ as three-components vectors,
\begin{equation}
|u\rangle
=
\left(
\begin{array}{c}
1 \\
0 \\
0 \\
\end{array}
\right),
\quad
|e\rangle
=
\left(
\begin{array}{c}
0 \\
1 \\
0 \\
\end{array}
\right),
\quad
|g\rangle
=
\left(
\begin{array}{c}
0 \\
0 \\
1 \\
\end{array}
\right).
\end{equation}
The Hamiltonian of the optical Bloch equations that controls the transition between $|u\rangle$ and $|e\rangle$
as the Rabi oscillation is given by
\begin{equation}
H_{\mbox{\scriptsize B}}
=
\hbar
\left(
\begin{array}{ccc}
0 & w(t) & 0 \\
w^{*}(t) & 0 & 0 \\
0 & 0 & 0 \\
\end{array}
\right),
\end{equation}
where
\begin{equation}
w(t)=(1/2)\Omega(t)\exp\{i[(E_{0}/\hbar)-\Delta]t\}.
\end{equation}
The time-dependent amplitude of the classical light is represented by $\Omega(t)$,
which is a complex number in general.
The Jaynes-Cummings interaction leading to the transition between $|g\rangle$ and $|e\rangle$ with the cavity field is given by the following Hamiltonian:
\begin{equation}
H_{\mbox{\scriptsize JC}}
=
\hbar
\left(
\begin{array}{ccc}
0 & 0 & 0 \\
0 & E_{0}/\hbar & ga \\
0 & g^{*}a^{\dagger} & (E_{0}/\hbar)-\Delta-\omega \\
\end{array}
\right)
+
\hbar\omega a^{\dagger}a,
\end{equation}
where $a$ and $a^{\dagger}$ denote the annihilation and creation operators of the cavity mode, respectively.
We assume the commutation relation $[a,a^{\dagger}]=1$.
The coupling constant $g$ is a complex number.
Thus, we can write down the total Hamiltonian of the single-photon source in the form,
\begin{eqnarray}
H
&=&
H_{\mbox{\scriptsize B}}+H_{\mbox{\scriptsize JC}} \nonumber \\
&=&
\hbar
\left(
\begin{array}{ccc}
0 & w(t) & 0 \\
w^{*}(t) & E_{0}/\hbar & ga \\
0 & g^{*}a^{\dagger} & (E_{0}/\hbar)-\Delta-\omega \\
\end{array}
\right)
+
\hbar\omega a^{\dagger}a.
\end{eqnarray}

The time evolution of the system obeys the Schr{\"o}dinger equation,
\begin{equation}
i\hbar\frac{\partial}{\partial t}\psi=H\psi,
\label{Schrodinger-equation-00}
\end{equation}
where the Hamiltonian $H$ is time-dependent because of $\Omega(t)$.
Here, we define the following unitary matrix:
\begin{equation}
U
=
\left(
\begin{array}{ccc}
\exp\{i[(E_{0}/\hbar)-\Delta]t\} & 0 & 0 \\
0 & 1 & 0 \\
0 & 0 & 1 \\
\end{array}
\right).
\end{equation}
Using the unitary matrix $U$,
we rewrite the wave function $\psi$ in Eq.~(\ref{Schrodinger-equation-00}) as follows:
\begin{equation}
\tilde{\psi}=U^{\dagger}\psi.
\end{equation}
Then, Eq.~(\ref{Schrodinger-equation-00}) changes into
\begin{equation}
i\hbar\frac{\partial}{\partial t}\tilde{\psi}=\tilde{H}\tilde{\psi},
\label{Schrodinger-equation-01}
\end{equation}
where the new Hamiltonian $\tilde{H}$ is given by
\begin{eqnarray}
\tilde{H}
&=&
-i\hbar U^{\dagger}\frac{\partial U}{\partial t}+U^{\dagger}HU \nonumber \\
&=&
E_{0}-\hbar\Delta
+
\hbar
\left(
\begin{array}{ccc}
0 & \Omega(t)/2 & 0 \\
\Omega^{*}(t)/2 & \Delta & ga \\
0 & g^{*}a^{\dagger} & -\omega \\
\end{array}
\right)
+
\hbar\omega a^{\dagger}a.
\label{Hamiltonian-02}
\end{eqnarray}
From now on, we neglect the constant term $(E_{0}-\hbar\Delta)$ in the right-hand side of Eq.~(\ref{Hamiltonian-02}).

We can divide the Hamiltonian given by Eq.~(\ref{Hamiltonian-02}) as
\begin{equation}
\tilde{H}=\tilde{H}_{0}+\tilde{H}_{1},
\label{Hamiltonian-03-a}
\end{equation}
\begin{equation}
\tilde{H}_{0}
=
\hbar
\left(
\begin{array}{ccc}
0 & 0 & 0 \\
0 & 0 & 0 \\
0 & 0 & -\omega \\
\end{array}
\right)
+
\hbar\omega a^{\dagger}a,
\label{Hamiltonian-03-b}
\end{equation}
\begin{equation}
\tilde{H}_{1}
=
\hbar
\left(
\begin{array}{ccc}
0 & \Omega(t)/2 & 0 \\
\Omega^{*}(t)/2 & \Delta & ga \\
0 & g^{*}a^{\dagger} & 0 \\
\end{array}
\right).
\label{Hamiltonian-03-c}
\end{equation}
Moreover, $\tilde{H}_{0}$ and $\tilde{H}_{1}$ satisfy the following commutation relation,
\begin{equation}
[\tilde{H}_{0},\tilde{H}_{1}]=0.
\label{commutation-relation-interaction-picture}
\end{equation}

Because we can diagonalize $\tilde{H}_{0}$ with ease,
we adopt the following interaction picture:
\begin{equation}
\tilde{\psi}_{\mbox{\scriptsize I}}(t)
=
\exp(i\tilde{H}_{0}t/\hbar)\tilde{\psi}(t),
\label{interaction-picture-0}
\end{equation}
where we assume
$\tilde{\psi}_{\mbox{\scriptsize I}}(0)=\tilde{\psi}(0)$.
Then, from Eqs.~(\ref{Schrodinger-equation-01}), (\ref{Hamiltonian-03-a}),
(\ref{commutation-relation-interaction-picture}), and (\ref{interaction-picture-0}),
we can describe the equation that $\tilde{\psi}_{\mbox{\scriptsize I}}(t)$ satisfies in the form,
\begin{equation}
i\hbar\frac{\partial}{\partial t}\tilde{\psi}_{\mbox{\scriptsize I}}(t)
=
\tilde{H}_{1}\tilde{\psi}_{\mbox{\scriptsize I}}(t).
\end{equation}
Thus, from now on, we regard $\tilde{H}_{1}$ as the Hamiltonian of the system.

As a result of the above discussion,
we can write down the Hamiltonian in the final form,
\begin{equation}
H
=
\left(
\begin{array}{ccc}
0 & \Omega(t)/2 & 0 \\
\Omega(t)/2 & \Delta & ga \\
0 & ga^{\dagger} & 0 \\
\end{array}
\right),
\label{Hamiltonian-04}
\end{equation}
where we rewrite the Hamiltonian of the interaction picture $\tilde{H}_{1}$ as $H$.
We put $\hbar=1$
and assume that both the amplitude of the classical light $\Omega(t)$ and the coupling constant of the Jaynes-Cummings interaction $g$ are real numbers.
The basis vectors of the Hilbert space,
where the Hamiltonian in Eq.~(\ref{Hamiltonian-04}) is defined,
are given by
$\{|i,n\rangle:i\in\{u,e,g\},n\in\{0,1,2,...\}\}$.
The index $n=0, 1, 2, ...$ denotes the number of photons in the cavity mode.

Because the Hamiltonian given by Eq.~(\ref{Hamiltonian-04}) is time-dependent,
we have to solve the time-dependent Schr{\"o}dinger equation that involves the first derivative
with respect to the time variable for pursuing the time evolution of the system.
However, we neglect these matters for a while and concentrate on obtaining eigenvalues and eigenvectors
of the Hamiltonian.

First of all, we consider which basis vectors are used to construct a superposition for the eigenvectors of the Hamiltonian $H$.
We pay attention to the following facts:
\begin{eqnarray}
H|u,n\rangle
&=&
[\Omega(t)/2]|e,n\rangle, \nonumber \\
H|e,n\rangle
&=&
[\Omega(t)/2]|u,n\rangle+\Delta|e,n\rangle+g\sqrt{n+1}|g,n+1\rangle, \nonumber \\
H|g,n\rangle
&=&
g\sqrt{n}|e,n-1\rangle.
\label{Hamiltonian-basis-vactors-relation-0}
\end{eqnarray}
Because of Eq.~(\ref{Hamiltonian-basis-vactors-relation-0}),
we can assume that the eigenvalue and its corresponding eigenvector are given in the form,
\begin{equation}
H|\omega\rangle=\omega|\omega\rangle,
\label{Hamiltonian-eigenvalue-relation-0}
\end{equation}
\begin{equation}
|\omega\rangle
=
c_{0}^{(n)}|u,n\rangle
+
c_{1}^{(n)}|e,n\rangle
+
c_{2}^{(n)}|g,n+1\rangle
\quad
\mbox{for $n=0,1,2,...$}.
\label{eigenvectors-form-0}
\end{equation}

Next, from Eqs.~(\ref{Hamiltonian-basis-vactors-relation-0}), (\ref{Hamiltonian-eigenvalue-relation-0}), and (\ref{eigenvectors-form-0}),
we obtain the eigenvalues for $n\in\{0,1,2,...\}$,
\begin{equation}
\omega=\omega_{0},\omega_{\pm},
\end{equation}
\begin{eqnarray}
\omega_{0}&=&0, \nonumber \\
\omega_{\pm}&=&\frac{1}{2}[\Delta\pm\sqrt{\Omega^{2}(t)+4g^{2}(n+1)+\Delta^{2}}].
\label{eigenvalues-form-0}
\end{eqnarray}
Their corresponding eigenvectors are given by
\begin{eqnarray}
|\omega_{0}\rangle
&=&
\left(
\begin{array}{c}
\cos\Theta_{n} \\
0 \\
-\sin\Theta_{n}
\end{array}
\right), \nonumber \\
|\omega_{+}\rangle
&=&
\left(
\begin{array}{c}
\cos\Phi_{n}\sin\Theta_{n} \\
-\sin\Phi_{n} \\
\cos\Phi_{n}\cos\Theta_{n} \\
\end{array}
\right), \nonumber \\
|\omega_{-}\rangle
&=&
\left(
\begin{array}{c}
\sin\Phi_{n}\sin\Theta_{n} \\
\cos\Phi_{n} \\
\sin\Phi_{n}\cos\Theta_{n} \\
\end{array}
\right),
\label{eigencectors-01}
\end{eqnarray}
where
\begin{equation}
|u,n\rangle
=
\left(
\begin{array}{c}
1 \\
0 \\
0 \\
\end{array}
\right),
\quad
|e,n\rangle
=
\left(
\begin{array}{c}
0 \\
1 \\
0 \\
\end{array}
\right),
\quad
|g,n+1\rangle
=
\left(
\begin{array}{c}
0 \\
0 \\
1 \\
\end{array}
\right),
\end{equation}
\begin{eqnarray}
\tan\Theta_{n}
&=&
\Omega(t)/(2g\sqrt{n+1}), \nonumber \\
\tan\Phi_{n}
&=&
\sqrt{4g^{2}(n+1)+\Omega^{2}(t)}/[\sqrt{4g^{2}(n+1)+\Omega^{2}(t)+\Delta^{2}}-\Delta].
\label{definition-Theta-Phi-0}
\end{eqnarray}

Here, we concentrate on the eigenvector whose corresponding eigenvalue is given by $\omega_{0}$ for $n=0$.
We write down it as follows:
\begin{equation}
|\omega_{0}\rangle
=
\left(
\begin{array}{c}
\cos\Theta_{0} \\
0 \\
-\sin\Theta_{0}\\
\end{array}
\right)
=
\cos\Theta_{0}|u,0\rangle-\sin\Theta_{0}|g,1\rangle,
\label{eigenvector-0}
\end{equation}
where
\begin{equation}
\tan\Theta_{0}
=
\frac{\Omega(t)}{2g}.
\label{eigenvector-1}
\end{equation}
Utilizing this eigenvector, we can realize the single-photon emitter.
We explain how to make use of it in the following.

First of all,
we put an initial state $|u,0\rangle$ for the atom-cavity system and give the classical light as $\Omega(t)=0$.
Next, we let the amplitude of the classical light $\Omega(t)$ increase sufficiently slowly as time passes.
If the time derivative of $\Omega(t)$ is small enough,
we can regard the time evolution of the wave function as an adiabatic process,
so that the wave function evolves with keeping its superposition in the form of the eigenvector $|\omega_{0}\rangle$
given by Eq.~(\ref{eigenvector-0}).
After much time has passed and $\Omega(t)$ has grown large enough,
the system changes into the state $|g,1\rangle$.

The time derivative of $\Omega(t)$ has to satisfy the adiabaticity constraint \cite{Messiah1961},
\begin{equation}
|\langle \omega_{\pm}|\frac{d}{dt}|\omega_{0}\rangle|\ll|\omega_{0}-\omega_{\pm}|.
\label{adiabatic-condition-0}
\end{equation}
On substitution from Eq.~(\ref{eigencectors-01}),
Eq.~(\ref{adiabatic-condition-0}) becomes
\begin{eqnarray}
|\dot{\Theta}_{0}\cos\Phi_{0}|&\ll&|\omega_{+}|, \nonumber \\
|\dot{\Theta}_{0}\sin\Phi_{0}|&\ll&|\omega_{-}|.
\end{eqnarray}
Thus, we obtain
\begin{equation}
|\dot{\Theta}_{0}|\ll|\omega_{\pm}|.
\label{adiabaticity-constraint-0}
\end{equation}

In the above adiabatic process, the state of the system always maintains the superposition of $|u,0\rangle$ and $|g,1\rangle$
and the system is prevented from reaching the state $|e,0\rangle$.
Thus, the system avoids the spontaneous emission of the classical light from the state $|e,0\rangle$
during the long-term adiabatic evolution.
This is the reason why we can expect the single-photon source to work in a quite stable manner.

If the system arrives at the state $|g,1\rangle$ with a high probability via the adiabatic process,
the single photon in the cavity mode is emitted through the imperfect mirror of the cavity.
Adjusting the decay rate of the cavity loss,
we can set proper length of the lifetime of $|g,1\rangle$.

\section{\label{section-trigger-pulse}The classical trigger pulse and the adiabaticity constraint}
In this section, we define the classical trigger light and derive an explicit form of the adiabaticity constraint.

In the present paper,
we give the amplitude of the classical light that induces the transition between the atomic states $|u\rangle$ and $|e\rangle$
in the Gaussian form,
\begin{equation}
\Omega(t)=\Omega_{0}\exp[-(t/T)^{2}],
\label{trigger-pulse-definition-0}
\end{equation}
where $\Omega_{0}$ and $T$ denote the characteristic amplitude and time, respectively.
We define full width at half maximum of $\Omega(t)$ as $2\tau$.
Then, $\tau$ is given by
\begin{equation}
\tau=\sqrt{\ln 2}T.
\end{equation}

Here, we derive an explicit form of the adiabaticity constraint.
From Eq.~(\ref{eigenvector-1}), we obtain
\begin{eqnarray}
\dot{\Theta}_{0}
&=&\cos^{2}\Theta_{0}\frac{\dot{\Omega}(t)}{2g} \nonumber \\
&=&\frac{2g\dot{\Omega}(t)}{4g^{2}+\Omega(t)^{2}}.
\end{eqnarray}
From Eq.~(\ref{eigenvalues-form-0}), putting $\Delta=0$ for the sake of simplicity, we obtain
\begin{equation}
|\omega_{\pm}|=\frac{1}{2}\sqrt{\Omega(t)^{2}+4g^{2}}.
\end{equation}
Thus, we can write down the adiabaticity constraint given by Eq.~(\ref{adiabaticity-constraint-0}) as
\begin{equation}
|\dot{\Omega}(t)|\ll\frac{1}{4g}(4g^{2}+\Omega(t)^{2})^{3/2}.
\end{equation}

Here, we assume $t=-\tau$, the half width at half maximum,
and the time derivative of the amplitude is given by
\begin{equation}
\dot{\Omega}(-\tau)
=
\frac{1}{T}\sqrt{\ln 2}\Omega_{0}.
\end{equation}
Thus, the adiabaticity constraint is expressed in the form,
\begin{equation}
T\gg\frac{4\sqrt{\ln 2}g\Omega_{0}}{[4g^{2}+(\Omega_{0}^{2}/4)]^{3/2}}.
\label{adiabaticity-form-0}
\end{equation}
In particular, taking $\Omega_{0}=4g$, we attain a simple form of the adiabaticity constraint as follows:
\begin{equation}
Tg\gg\sqrt{\ln 2}/\sqrt{2}\simeq 0.588\mbox{\,}70.
\label{adiabaticity-form-1}
\end{equation}

\section{\label{section-master-equation}The master equation that describes the emission of a single photon in the cavity mode as the cavity loss}
In this section, we introduce the master equation that describes the transmission of the single photon through the imperfect mirror.

If the system reaches the state $|g,1\rangle$ via the adiabatic process,
it has to emit a single photon in the cavity mode to the outside of the optical cavity
for realizing the single-photon gun.
This implies that we need to let the single photon pass through the mirror.
To pursue the cavity loss induced by the imperfect mirror of the cavity,
we employ the following master equation:
\begin{equation}
\dot{\rho}(t)
=
-i[H(t),\rho(t)]
+
\gamma[a\rho(t)a^{\dagger}-\frac{1}{2}(a^{\dagger}a\rho(t)+\rho(t)a^{\dagger}a)],
\label{master-equation-0}
\end{equation}
where we write the Hamiltonian as $H(t)$ for emphasizing its time dependence.

We pay attention to the fact that the dynamics of the master equation~(\ref{master-equation-0}) is restricted
inside the four dimensional Hilbert space ${\cal H}_{4}=\{|u,0\rangle,|e,0\rangle,|g,1\rangle,|g,0\rangle\}$.
Thus, to solve the master equation (\ref{master-equation-0}) numerically,
we only have to consider ${\cal H}_{4}$.

The transition $|g,1\rangle\to|g,0\rangle$ emits the single photon in the cavity mode to the outside of the cavity.
By repetition, some photons contribute to the single-photon generation and the others cause leakage through the mirrors.
In Ref.~\cite{Keller2004a}, a rate $\gamma_{t}$ with $0\leq\gamma_{t}\leq\gamma$ was introduced
for only including the transmission of photons through the mirror
as the single-photon gun.
The authors of Ref.~\cite{Keller2004a} put $\gamma_{t}=0.9\mbox{\,}\gamma$.
The rate of emission from the cavity is given by
\begin{equation}
P(t)=\gamma_{t}\mbox{Tr}\{a^{\dagger}a\rho(t)\}.
\end{equation}

Because the dynamics of the system lies on the Hilbert space ${\cal H}_{4}$,
we can express the rate of the emission in the form,
\begin{equation}
P(t)=\gamma_{t}p(t),
\label{rate-of-the-emission-01}
\end{equation}
\begin{equation}
p(t)=\langle g,1|\rho(t)|g,1\rangle.
\label{rate-of-the-emission-02}
\end{equation}
The efficiency of single-photon generation is given by
\begin{equation}
\eta=\int_{-\infty}^{\infty}P(t)dt.
\label{rate-of-the-emission-03}
\end{equation}
Here, we pay attention to the fact that $\eta$ is a dimensionless quantity.
In the present paper, for the sake of simplicity,
we put $\gamma_{t}=\gamma$.

We let $\delta t$ denote full width at half maximum for $p(t)$.
That is to say, letting $t_{\mbox{\scriptsize max}}$ be the time when $p(t)$ becomes maximum,
$t_{-}$ be the time when $p(t)$ is equal to a half of its maximum value before $t_{\mbox{\scriptsize max}}$,
and
$t_{+}$ be the time when $p(t)$ is equal to a half of its maximum value after $t_{\mbox{\scriptsize max}}$,
we put $\delta t=t_{+}-t_{-}$.
We measure $t_{\mbox{\scriptsize max}}$, $t_{-}$, and $t_{+}$ from the peak of the classical trigger pulse.
The reason why we do not consider full width at half maximum for $P(t)$ but for $p(t)$ is
that the full width at half maximum for $P(t)$ is not continuous at $\gamma=0$.
For $\gamma\neq 0$, the full width at half maximum for $P(t)$ is equal to that for $p(t)$.
We can regard $\delta t$ as the fluctuation of the duration of the emission of the single photon.
In addition, we can regard $t_{\mbox{\scriptsize max}}$ as the time when the single photon is emitted.
In the current paper, we examine the dependence of $\eta$, $\delta t$, and $t_{\mbox{\scriptsize max}}$
on $T$, $\gamma$, $g$, and $\Omega_{0}$.

\begin{figure}
\begin{center}
\includegraphics{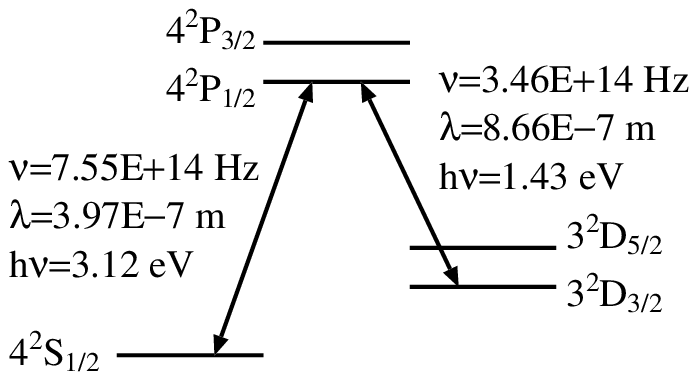}
\end{center}
\caption{Energy levels of ${}^{40}\mbox{Ca}^{+}$.}
\label{Figure03}
\end{figure}

Here, we give numerical parameters for solving the master equation actually.
In Ref.~\cite{Keller2004a}, the on-demand single-photon source was experimentally realized using a calcium ion in a cavity,
whose simplified level scheme is shown in Fig.~\ref{Figure03}.
The amplitude of the classical trigger pulse was given by $\Omega_{0}=0.11\times 22\times 2\pi$~MHz in Ref.~\cite{Keller2004a}.
In numerical calculations of the current paper, we adopt this quantity.
As a typical value of the coupling constant for the Jaynes-Cummings interaction,
we choose $g=\Omega_{0}/4$ for example.
For the sake of simplicity, we let the detuning be given by $\Delta=0$.
We take $T=5.0\times 10^{-5}$~s for the characteristic time of the trigger pulse given by Eq.~(\ref{trigger-pulse-definition-0}) for instance.
Then, we obtain $Tg\simeq 190.07\gg 0.588\mbox{\,}70$ and the adiabaticity constraint (\ref{adiabaticity-form-1}) holds.
We assume that the rate of the transmission of the photon through the imperfect mirror is in the range of $0\leq\gamma\leq 0.4$~MHz.

In the current paper, we give physical quantities in two significant figures.
However, to keep their accuracy, we carry out numerical calculations, for example, solving the master equation,
with five significant figures throughout the present paper.

\section{\label{section-numerical-calculation-populations}Time evolution of the population for each state of the system}
\begin{figure}
\begin{center}
\includegraphics{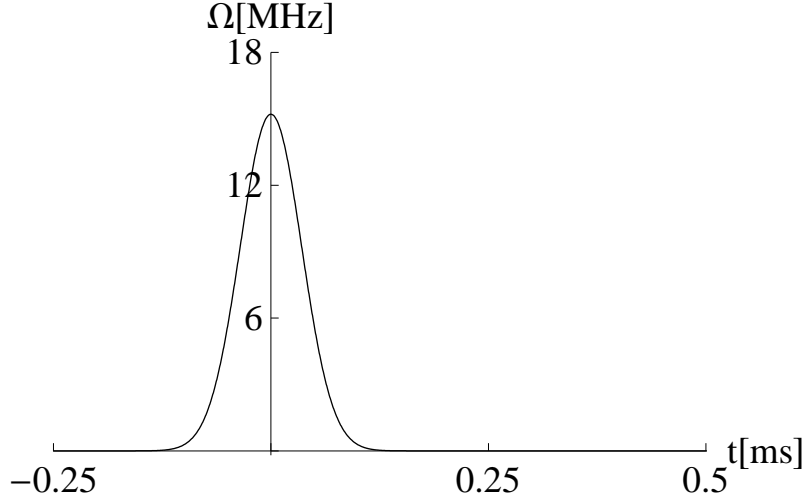}
\end{center}
\caption{Time evolution of the classical trigger pulse $\Omega(t)$ for
$\Omega_{0}=0.11\times 22\times 2\pi$~MHz
and $T=5.0\times 10^{-5}$~s in Eq.~(\ref{trigger-pulse-definition-0}).}
\label{Figure04}
\end{figure}

From now on, in the succeeding four sections,
we report numerical results obtained by solving the master equation given by Eq.~(\ref{master-equation-0}).
How to solve the master equation numerically is explained in Appendix~\ref{section-numerical-calculation-master-equation}.
In this section, we examine the time evolution of the populations of the states
$|u,0\rangle$, $|e,0\rangle$, $|g,1\rangle$, and $|g,0\rangle$.
We assume that the trigger pulse is given by
$\Omega_{0}=0.11\times 22\times 2\pi \mbox{ MHz}\simeq 1.5205\times 10^{7} \mbox{ Hz}$
and $T=5.0\times 10^{-5}$~s in Eq.~(\ref{trigger-pulse-definition-0}).
Figure~\ref{Figure04} shows the time evolution of $\Omega(t)$.

\begin{figure}
\begin{center}
\includegraphics{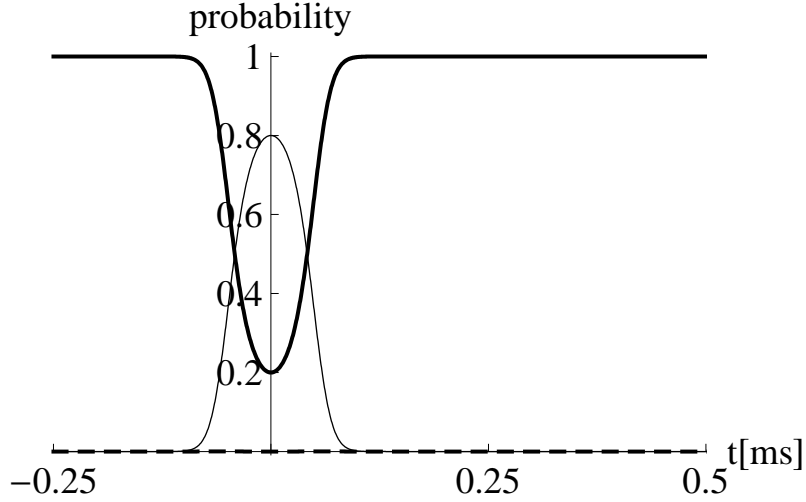}
\end{center}
\caption{Time evolution of populations of the states
$|u,0\rangle$, $|e,0\rangle$, $|g,1\rangle$, and $|g,0\rangle$
for $\Omega_{0}=0.11\times 22\times 2\pi$~MHz,
$T=5.0\times 10^{-5}$~s,
$g=\Omega_{0}/4$, and $\gamma=0$.
A thick solid curve, a thick dashed curve, a thin solid curve, and a thin dashed curve
represent the populations of $|u,0\rangle$, $|e,0\rangle$, $|g,1\rangle$, and $|g,0\rangle$, respectively.
Looking at the graph for $|e,0\rangle$,
we notice that the adiabatic evolution is realized.
The population of $|u,0\rangle$ at $t=0$ is nearly equal to $1/5$.}
\label{Figure05}
\end{figure}

Figure~\ref{Figure05} gives the time evolution of the populations for the states with
$g=\Omega_{0}/4\simeq 3.8013\times 10^{6}$~Hz
and
$\gamma=0$.
Solving the master equation numerically to obtain results in Fig.~\ref{Figure05},
we start calculations from the time
$t=-5T=-2.5\times 10^{-4}$~s
with putting the initial state $|u,0\rangle$.
Looking at Fig.~\ref{Figure05},
we notice the population of $|e,0\rangle$ be always nearly equal to zero and the adiabatic process be realized.
The population of $|g,1\rangle$, namely $p(t)$,
takes the maximum value at $t=0$, so that $t_{\mbox{\scriptsize max}}=0$.
The full width at half maximum for $p(t)$ is given by $\delta t\simeq 9.4653\times 10^{-5}$~s.
Because $\gamma=0$, the transition from $|g,1\rangle$ to $|g,0\rangle$ is prevented
and we can confirm that the population of $|g,0\rangle$ is always equal to zero.

Because of the adiabatic evolution with Eqs.~(\ref{eigenvector-0}) and (\ref{eigenvector-1}),
the population of $|g,1\rangle$ approximates to
\begin{eqnarray}
p(t)
&\simeq&
\sin^{2}\Theta_{0} \nonumber \\
&=&
\frac{[\Omega(t)/(2g)]^{2}}{1+[\Omega(t)/(2g)]^{2}}.
\label{population-p-1}
\end{eqnarray}
From $\Omega(0)=\Omega_{0}$ and $g=\Omega_{0}/4$,
we obtain $p(0)\simeq 4/5$.
Thus, the population of $|u,0\rangle$ at $t=0$ is nearly equal to $1/5$.
We can recognize this fact in Fig.~\ref{Figure05}.
This observation implies that we have checked the programming code for solving the master equation and we can convince ourselves of the numerical results obtained.

\begin{figure}
\begin{center}
\includegraphics{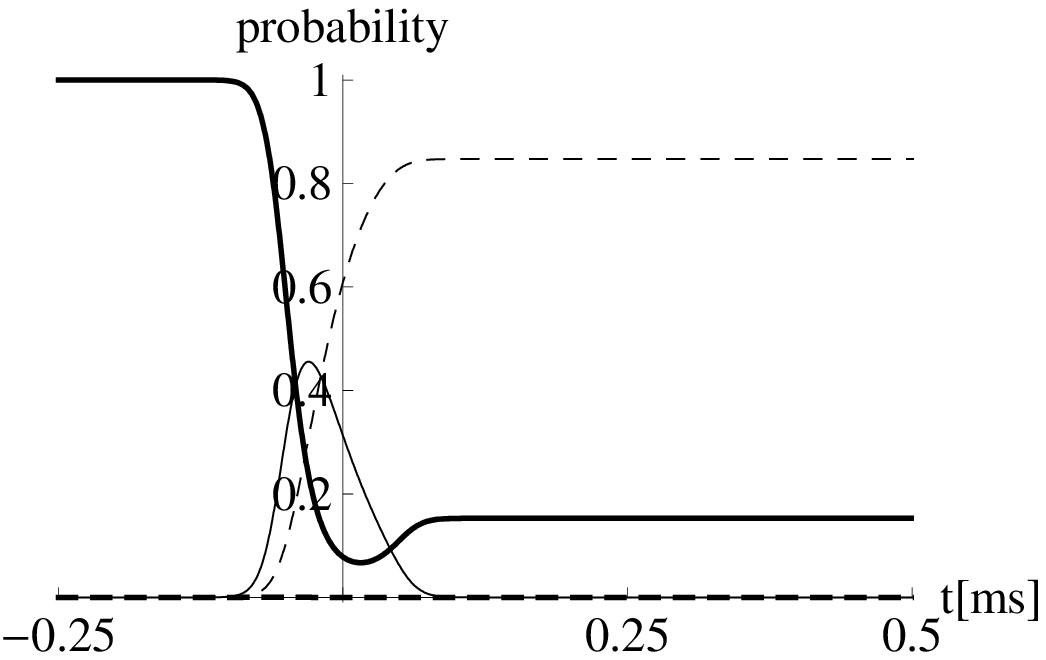}
\end{center}
\caption{Time evolution of populations of the states
$|u,0\rangle$, $|e,0\rangle$, $|g,1\rangle$, and $|g,0\rangle$
for $\Omega_{0}=0.11\times 22\times 2\pi$~MHz,
$T=5.0\times 10^{-5}$~s,
$g=\Omega_{0}/4$, and $\gamma=2.5\times 10^{4}$~Hz.
A thick solid curve, a thick dashed curve, a thin solid curve, and a thin dashed curve
represent the populations of $|u,0\rangle$, $|e,0\rangle$, $|g,1\rangle$, and $|g,0\rangle$, respectively.
Turning our eyes to the graph for $|e,0\rangle$,
we notice that the adiabatic evolution is realized.
A shape of $p(t)$, the population of $|g,1\rangle$,
is not symmetric with respect to a vertical axis $t=t_{\mbox{\scriptsize max}}$.}
\label{Figure06}
\end{figure}

Figure~\ref{Figure06} shows the time evolution of the populations of the states with putting $\gamma=2.5\times 10^{4}$~Hz.
Values of the parameters $\Omega_{0}$, $T$, and $g$ are equal to those in Fig.~\ref{Figure05}.
In Fig.~\ref{Figure06}, we can observe the population of $|e,0\rangle$ be always nearly equal to zero, as well.
Thus, we can confirm that the adiabatic evolution occurs in Fig.~\ref{Figure06}.
In Fig.~\ref{Figure06}, because of $t_{\mbox{\scriptsize max}}\simeq -3.0173\times 10^{-5}$~s,
we can consider that the single photon is emitted earlier concerning the peak of the trigger pulse.
Moreover, a shape of $p(t)$, the population of $|g,1\rangle$,
is not symmetric with respect to a vertical axis $t=t_{\mbox{\scriptsize max}}$.
This is because the transition from $|g,1\rangle$ to $|g,0\rangle$ happens with the decay rate $\gamma(\neq 0)$.

In Fig.~\ref{Figure06}, the full width at half maximum for $p(t)$ is given by $\delta t\simeq 7.0393\times 10^{-5}$~s.
The value of $p(t)$ becomes nearly equal to zero for $t\geq 1.2\times 10^{-4}$~s.
The populations of $|u,0\rangle$ and $|g,0\rangle$ come to rest on values $0.152\mbox{\,}80$ and $0.847\mbox{\,}20$ around, respectively.
The efficiency of the single-photon generation is given by $\eta\simeq 0.847\mbox{\,}20$.

\section{\label{section-efficiency}The efficiency of the single-photon generation}
\begin{figure}
\begin{center}
\includegraphics{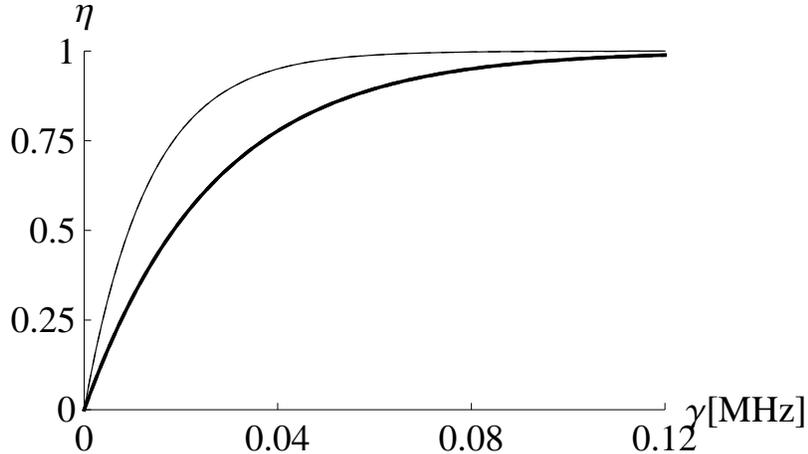}
\end{center}
\caption{Graphs of the efficiency of the single-photon generation $\eta$ shown as functions of the decay rate $\gamma$
with putting $\Omega_{0}=0.11\times 22\times 2\pi$~MHz and $g=\Omega_{0}/4$.
A thick solid curve and a thin solid curve represent cases with $T=2.5\times 10^{-5}$~s and $T=5.0\times 10^{-5}$~s, respectively.
In addition, graphs of the function given by Eqs.~(\ref{eta-approximate-plot-0}) and (\ref{eta-approximate-plot-1})
are plotted with a thick dashed curve and a thin dashed curve for $T=2.5\times 10^{-5}$~s and $T=5.0\times 10^{-5}$~s, respectively.
Because differences between numerical results obtained by the master equation and the calculated values derived
from Eqs.~(\ref{eta-approximate-plot-0}) and (\ref{eta-approximate-plot-1}) are too small,
we cannot distinguish dashed curves from solid curves.
This fact is evidence that approximation with the function given by Eqs.~(\ref{eta-approximate-plot-0}) and (\ref{eta-approximate-plot-1})
is very precise.}
\label{Figure07}
\end{figure}

In this section, we examine the efficiency of the single-photon generation numerically.
In Fig.~\ref{Figure07}, graphs of the efficiency $\eta$ are plotted as functions of the decay rate $\gamma$.
We put the physical parameters
$\Omega_{0}=0.11\times 22\times 2\pi$~MHz and $g=\Omega_{0}/4$ in Fig.~\ref{Figure07},
as they are given in Figs.~\ref{Figure05} and \ref{Figure06}.
We can reconstruct the results of the numerical calculations with the following function approximately:
\begin{equation}
\eta\simeq 1-\exp(-aT\gamma),
\label{eta-approximate-plot-0}
\end{equation}
\begin{equation}
a\simeq 1.5029.
\label{eta-approximate-plot-1}
\end{equation}
For the case with $T=2.5\times 10^{-5}$~s in Fig.~\ref{Figure07},
a difference between numerical results obtained by the master equation and calculated values derived
from Eqs.~(\ref{eta-approximate-plot-0}) and (\ref{eta-approximate-plot-1}) is equal to or less than $8.7244\times 10^{-7}$.
Hence, Eqs.~(\ref{eta-approximate-plot-0}) and (\ref{eta-approximate-plot-1}) give a close approximation to numerical data obtained by solving the master equation.
From numerical calculations with putting various values for $\Omega_{0}$, $g$, and $T$,
we observe the following result.
Although we let both values of the characteristic amplitude and time,
$\Omega_{0}$ and $T$,
vary at random,
Eqs.~(\ref{eta-approximate-plot-0}) and (\ref{eta-approximate-plot-1}) hold with $g=\Omega_{0}/4$.
Thus, we can conclude that the parameter $a$ in Eq.~(\ref{eta-approximate-plot-0}) has to depend only on $g/\Omega_{0}$.

\begin{figure}
\begin{center}
\includegraphics{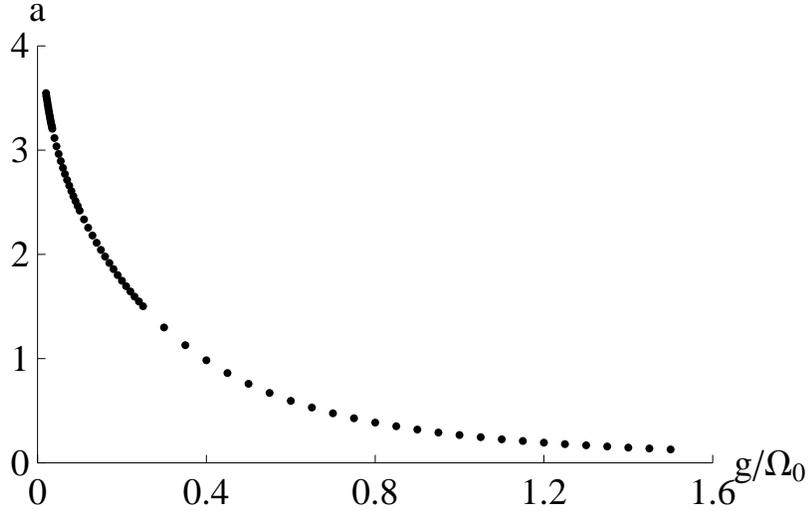}
\end{center}
\caption{Variation of the parameter $a$ with $g/\Omega_{0}$.
In the sequence of small black circles, the value of $g/\Omega_{0}$ is in the range of $0.02\leq g/\Omega_{0}\leq 1.5$.
If we take parameters $\Omega_{0}=0.11\times 22\times 2\pi$~MHz and $g/\Omega_{0}=0.02$,
we obtain
$4\sqrt{\ln 2}g\Omega_{0}/[4g^{2}+(\Omega_{0}^{2}/4)]^{3/2}\simeq 3.4709\times 10^{-8}$~s.
Because of $T=5.0\times 10^{-5}$~s, the adiabaticity constraint (\ref{adiabaticity-form-0}) is satisfied.}
\label{Figure08}
\end{figure}

\begin{figure}
\begin{center}
\includegraphics{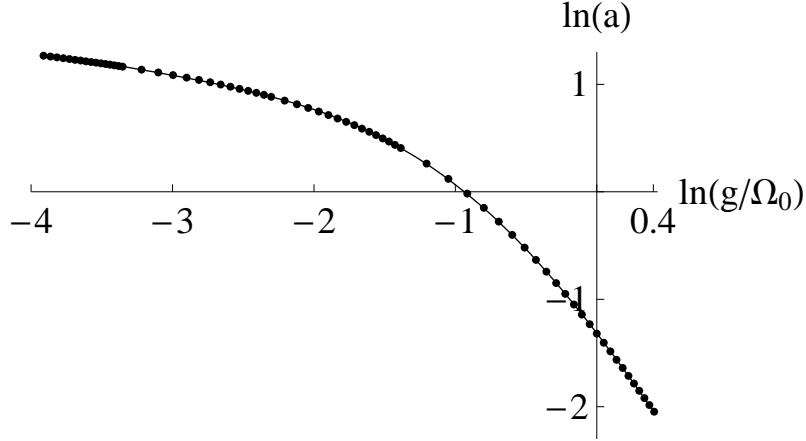}
\end{center}
\caption{Variation of $\ln a$ with $\ln(g/\Omega_{0})$.
Each small black circle is obtained by solving the master equation numerically.
A thin solid curve represents the fifth degree polynomial
given by Eqs.~(\ref{a_g_Omega0_fitting_0}), (\ref{a_g_Omega0_fitting_1}), (\ref{a_g_Omega0_fitting_2}), and (\ref{a_g_Omega0_fitting_3})
which has the best fit to the sequence of the small black circles.}
\label{Figure09}
\end{figure}

Figure~\ref{Figure08} shows variation of the parameter $a$ with $g/\Omega_{0}$.
Figure~\ref{Figure09} shows variation of $\ln a$ with $\ln(g/\Omega_{0})$.
In both Figs.~\ref{Figure08} and \ref{Figure09},
small black circles represent numerical results obtained by solving the master equation.
To obtain each black circle, we carry out the following task.
First, we put $T=5.0\times 10^{-5}$~s and $\Omega_{0}=0.11\times 22\times 2\pi$~MHz
and set $g/\Omega_{0}$ to a specific value.
Second, we obtain variation of the efficiency $\eta$ with $\gamma$ for $0\leq\gamma\leq 3.6\times 10^{5}$~Hz
by solving the master equation numerically.
Third, we fit the function given by Eq.~(\ref{eta-approximate-plot-0})
to numerical data points of the variation of $\eta$, so that we obtain the parameter $a$.
We can fit the following fifth degree polynomial to the sequence of the small black circles:
\begin{equation}
y=b_{0}+b_{1}x+b_{2}x^{2}+b_{3}x^{3}+b_{4}x^{4}+b_{5}x^{5},
\label{a_g_Omega0_fitting_0}
\end{equation}
\begin{equation}
x=\ln(g/\Omega_{0}),
\label{a_g_Omega0_fitting_1}
\end{equation}
\begin{equation}
y=\ln a,
\label{a_g_Omega0_fitting_2}
\end{equation}
\begin{eqnarray}
b_{0}&\simeq&-1.3173,
\quad
b_{1}\simeq -1.7179,
\quad
b_{2}\simeq -0.263\mbox{\,}29, \nonumber \\
b_{3}&\simeq&
{0.114\mbox{\,}87},
\quad
b_{4}\simeq 0.048\mbox{\,}967,
\quad
b_{4}\simeq 0.005\mbox{\,}135\mbox{\,}8.
\label{a_g_Omega0_fitting_3}
\end{eqnarray}
In Fig.~\ref{Figure09}, we plot the polynomial
given by Eqs.~(\ref{a_g_Omega0_fitting_0}), (\ref{a_g_Omega0_fitting_1}), (\ref{a_g_Omega0_fitting_2}), and (\ref{a_g_Omega0_fitting_3})
with a thin solid curve.

Looking at Figs.~\ref{Figure08}, \ref{Figure09}, and Eq.~(\ref{eta-approximate-plot-0}),
we notice that the parameter $a$ increases and the efficiency $\eta$ becomes easy to attain unity as $g/\Omega_{0}$ declines.
This implies the following.
The efficiency rises if the transition caused by the optical Bloch equations is superior to the transition induced by the Jaynes-Cummings interaction.
In other words, as we let the pump intensity $\Omega_{0}$ increase with fixing the coupling constant of the Jaynes-Cummings interaction $g$ to a specific value,
the efficiency of the emission of the single photon approaches unity.

From a different viewpoint,
we consider the above facts.
As mentioned in the previous paragraph, Figs.~\ref{Figure08} and \ref{Figure09} show that the parameter $a$ increases with smaller $g/\Omega_{0}$.
Thus, remembering Eq.~(\ref{eta-approximate-plot-0}),
we become aware that the efficiency of the photon generation $\eta$ approaches unity as the coupling constant $g$ diminishes.
However, in general, stronger coupling between the atom and the cavity mode increases the photon generation efficiency.
These two facts seem to contradict each other.

In point of fact, for the photon generation scheme the current paper deals with,
$\eta$ becomes larger as $g$ gets smaller.
We can find an indication of this phenomenon in Eqs.~(\ref{eigenvector-0}), (\ref{eigenvector-1}), and (\ref{trigger-pulse-definition-0}).
According to Eq.~(\ref{eigenvector-1}),
if $\Omega_{0}/g$ increases,
the population of the state $|u,0\rangle$ is suppressed and that of $|g,1\rangle$ is enhanced.
In particular, looking at Eq.~(\ref{population-p-1}),
we clearly understand that the population of $|g,1\rangle$ attains unity as $\Omega_{0}/g$ becomes larger for $\gamma=0$ and $t=0$.
Thus, from Eqs.~(\ref{rate-of-the-emission-01}), (\ref{rate-of-the-emission-02}), and (\ref{rate-of-the-emission-03}),
for the scheme discussed in the current paper,
the efficiency of the photon generation $\eta$ is amplified by decrease of the coupling strength $g$.
This fact seems to be inconsistent with common sense about the general atom-cavity system.

The physical meaning of this discrepancy is as follows.
The coupling constant $g$ governs the transition between $|e,0\rangle$ and $|g,1\rangle$.
However, because of the adiabatic process,
the population of $|e,0\rangle$ is always nearly equal to zero and the transition between $|e,0\rangle$ and $|g,1\rangle$ is prohibited.
Hence, basically, a magnitude of $g$ has nothing to do with the photon generation.

However, because of the adiabatic process, the wave function $|\omega_{0}\rangle$ given by Eqs.~(\ref{eigenvector-0}) and (\ref{eigenvector-1})
evolves with keeping its superposition.
Thus, the population of $|g,1\rangle$ is enhanced with larger $\Omega_{0}/g$.
In other words, the efficiency of the photon generation attains unity as $g$ gets smaller.
Here, we have to pay attention to the adiabaticity constraint given by Eq.~(\ref{adiabaticity-form-0}).
According to Eq.~(\ref{adiabaticity-form-0}),
we cannot let the coupling constant $g$ be smaller freely.
For example, we cannot set $g=0$.
Hence, we conclude that the efficiency $\eta$ approaches unity as the coupling constant $g$ gets smaller as far as the adiabaticity constraint holds.

\section{\label{section-time-fluctuation}The fluctuation of the duration of the photon emission}
In this section, we numerically examine $\delta t$,
the fluctuation of the duration of the single-photon emission.
We put $\Omega_{0}=0.11\times 22\times 2\pi$~MHz throughout this section.

First, we consider $\delta t$ for $\gamma=0$.
Under the adiabatic approximation, letting $\gamma=0$,
we obtain the population of $|g,1\rangle$ as $p(t)$ in Eq.~(\ref{population-p-1}),
where $\Omega(t)$ is given by Eq.~(\ref{trigger-pulse-definition-0}).
Thus, solving an equation $p(t)=p(0)/2$ with Eqs.~(\ref{trigger-pulse-definition-0}) and (\ref{population-p-1}),
we obtain the full width at the half maximum,
\begin{equation}
\delta t=\{2\ln[2+(1/4)(\Omega_{0}/g)^{2}]\}^{1/2}T.
\label{half-time-T-0}
\end{equation}
Looking at Eq.~(\ref{half-time-T-0}),
we notice that $\delta t$ depends only on $T$ and $g/\Omega_{0}$
and it is a linear function with respect to $T$.
In particular, we obtain $\delta t=\sqrt{2\ln 6}T\simeq 1.8930\mbox{\,}T$ for $g/\Omega_{0}=1/4$.

\begin{figure}
\begin{center}
\includegraphics{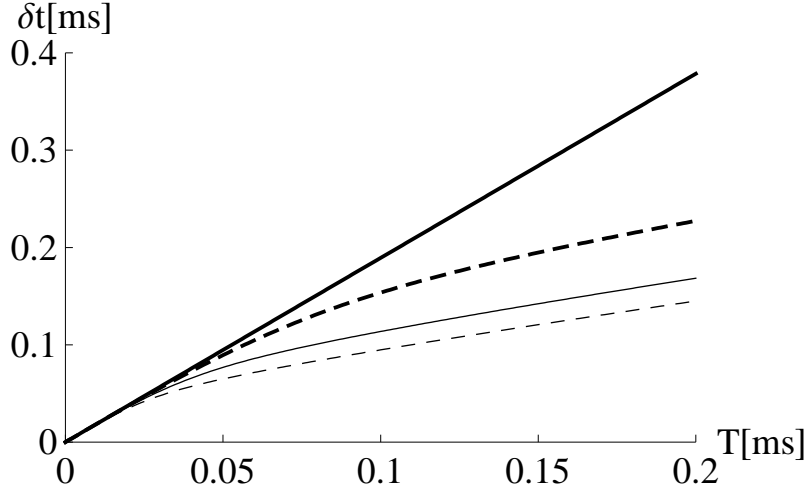}
\end{center}
\caption{The calculated fluctuation $\delta t$ versus $T$ with putting $g/\Omega_{0}=1/4$.
A thick solid line, a thick dashed curve, a thin solid curve, and a thin dashed curve represent the fluctuations
for $\gamma=0$, $\gamma=1.0\times 10^{4}$~Hz, $\gamma=2.0\times 10^{4}$~Hz, and $\gamma=3.0\times 10^{4}$~Hz,
respectively.
The thick solid line corresponds with $\delta t\simeq 1.8930\mbox{\,}T$,
namely Eq.~(\ref{half-time-T-0}), well.}
\label{Figure10}
\end{figure}

\begin{figure}
\begin{center}
\includegraphics{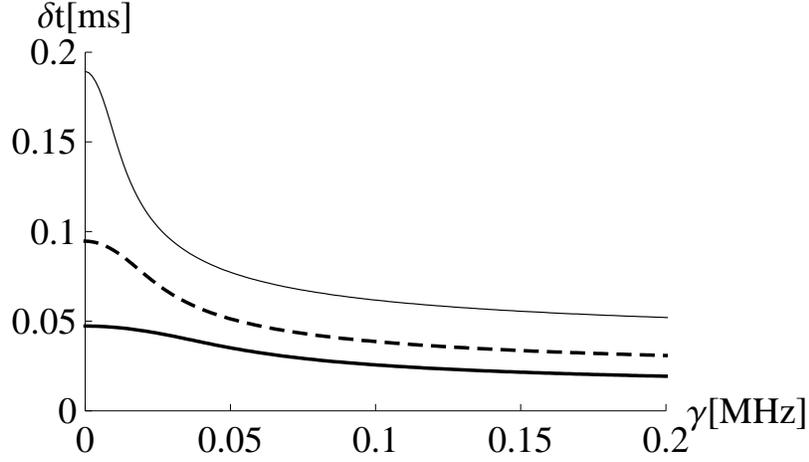}
\end{center}
\caption{Variation of the fluctuation $\delta t$ as a function of $\gamma$ with putting $g/\Omega_{0}=1/4$.
A thick solid curve, a thick dashed curve, and a thin solid curve represent the fluctuations
for $T=2.5\times 10^{-5}$~s, $T=5.0\times 10^{-5}$~s, and $T=1.0\times 10^{-4}$~s, respectively.
The fluctuation of the duration of the single-photon emission $\delta t$ decreases as $\gamma$ becomes larger.}
\label{Figure11}
\end{figure}

\begin{figure}
\begin{center}
\includegraphics{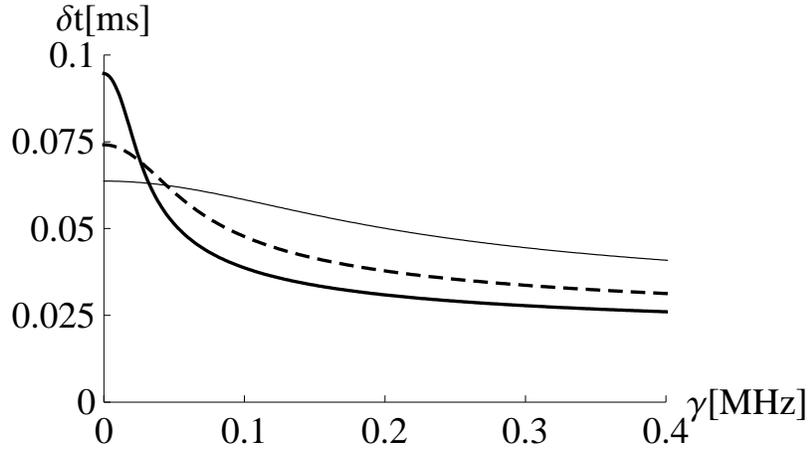}
\end{center}
\caption{Variation of the fluctuation $\delta t$ with $\gamma$ for $T=5.0\times 10^{-5}$~s.
A thick solid curve, a thick dashed curve, and a thin solid curve represent the fluctuations
for $g/\Omega_{0}=0.25$, $g/\Omega_{0}=0.5$, and $g/\Omega_{0}=1.0$, respectively.
Variation of $\delta t$ becomes more gradual as $g/\Omega_{0}$ increases.}
\label{Figure12}
\end{figure}

Figure~\ref{Figure10} shows calculated $\delta t$ versus $T$ with putting $g/\Omega_{0}=1/4$.
Turning our eyes to Fig.~\ref{Figure10}, we observe that $\delta t$ is not a linear function concerning $T$ for $\gamma\neq 0$.
Moreover, we understand that Eq.~(\ref{half-time-T-0}) gives the upper bound of $\delta t$.
Figure~\ref{Figure11} shows variation of $\delta t$ as a function of $\gamma$ with putting $g/\Omega_{0}=1/4$.
Looking at Fig.~\ref{Figure11}, we perceive that $\delta t$ decreases as $\gamma$ rises.
However, in Fig.~\ref{Figure11}, $\delta t$ declines slowly for $\gamma\geq 1.0\times 10^{5}$~Hz.
Figure~\ref{Figure12} shows variation of $\delta t$ with $\gamma$ for $T=5.0\times 10^{-5}$~s.
Turning our eyes to Fig.~\ref{Figure12}, we observe that the variation becomes more gradual as $g/\Omega_{0}$ increases.

From Eq.~(\ref{eta-approximate-plot-0}),
fixing $\gamma$ to a specific value,
we can let the efficiency $\eta$ approach unity by increasing $T$.
However, Eq.~(\ref{half-time-T-0}) tells us that elevation of $T$ lets $\delta t$ become larger.
Thus, there is a trade-off between the efficiency of generation of the single photon
and the fluctuation of the duration of the emission in respect of the value of $T$.

Figures~\ref{Figure08} and \ref{Figure09} tell us that the parameter $a$ increases as $g/\Omega_{0}$ becomes smaller.
Thus, from Eq.~(\ref{eta-approximate-plot-0}), the efficiency $\eta$ approaches unity as $g/\Omega_{0}$ decreases.
In contrast, letting $g/\Omega_{0}$ be smaller,
we obtain $\delta t$ getting larger from Eq.~(\ref{half-time-T-0}).
Hence, we find another trade-off between $\eta$ and $\delta t$ with respect to $g/\Omega_{0}$.

\section{\label{section-t-max}Variation of the time when $p(t)$ becomes maximum}
In this section, we numerically study the time when the single photon is emitted through the imperfect mirror of the cavity.
As shown in Fig.~\ref{Figure06},
the emission of the photon occurs earlier concerning the peak of the trigger pulse.
This is because the state $|g,1\rangle$ changes into the state $|g,0\rangle$ with the decay rate $\gamma$.
Throughout this section, we put $\Omega_{0}=0.11\times 22\times 2\pi$~MHz.

\begin{figure}
\begin{center}
\includegraphics{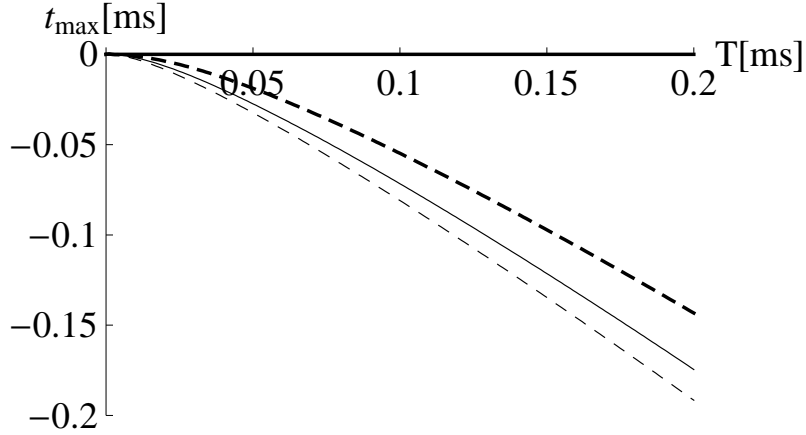}
\end{center}
\caption{Variation of $t_{\mbox{\scriptsize max}}$ with $T$ for $g/\Omega_{0}=1/4$.
A thick solid line, a thick dashed curve, a thin solid curve, and a thin dashed curve represent cases
with $\gamma=0$, $\gamma=1.0\times 10^{4}$~Hz, $\gamma=2.0\times 10^{4}$~Hz, and $\gamma=3.0\times 10^{4}$~Hz, respectively.
In these calculations, we put $g=\Omega_{0}/4\simeq 3.8013\times 10^{6}$~Hz.
Because of Eq.~(\ref{adiabaticity-form-1}), the adiabaticity constraint is given by $T\gg 1.5487\times 10^{-7}$~s.
Thus, the adiabaticity constraint is not fulfilled near the origin of the axes.
Looking at these graphs, we observe that $t_{\mbox{\scriptsize max}}$ changes position in the negative direction as $T$ and $\gamma$ become larger.}
\label{Figure13}
\end{figure}

\begin{figure}
\begin{center}
\includegraphics{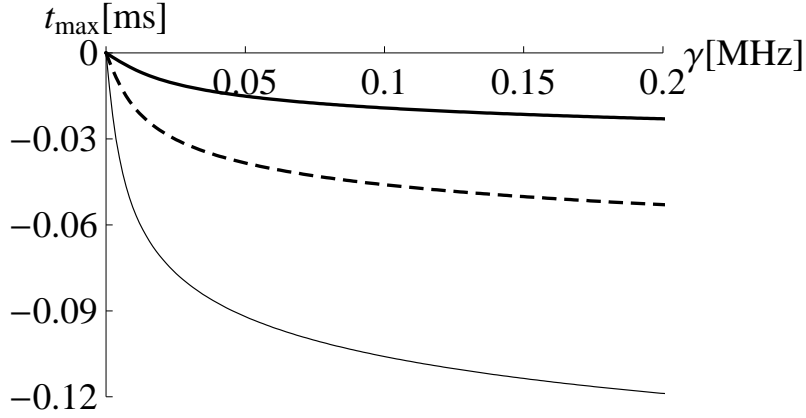}
\end{center}
\caption{Calculated $t_{\mbox{\scriptsize max}}$ versus $\gamma$ with putting $g/\Omega_{0}=1/4$.
A thick solid curve, a thick dashed curve, and a thin solid curve represent cases
with $T=2.5\times 10^{-5}$~s, $T=5.0\times 10^{-5}$~s, and $T=1.0\times 10^{-4}$~s, respectively.
Turning our eyes to these graphs, we observe that $t_{\mbox{\scriptsize max}}$ changes position in the negative direction as $T$ and $\gamma$ become larger.}
\label{Figure14}
\end{figure}

\begin{figure}
\begin{center}
\includegraphics{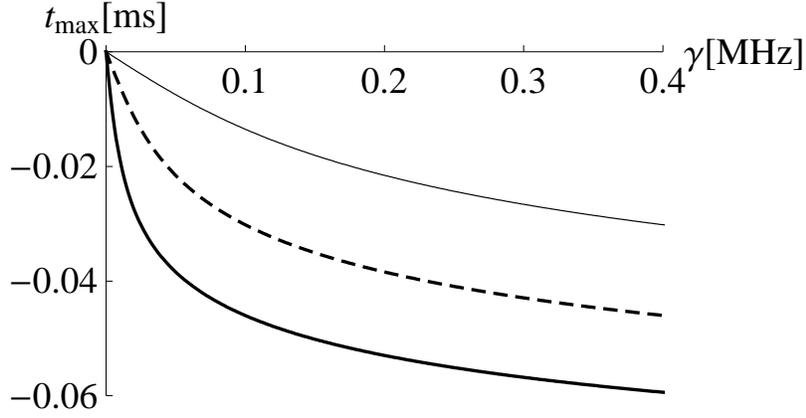}
\end{center}
\caption{Variation of $t_{\mbox{\scriptsize max}}$ as a function of $\gamma$ with putting $T=5.0\times 10^{-5}$~s.
A thick solid curve, a thick dashed curve, and a thin solid curve represent cases
with $g/\Omega_{0}=0.25$, $g/\Omega_{0}=0.5$, and $g/\Omega_{0}=1.0$, respectively.
Looking at these graphs, we perceive that $t_{\mbox{\scriptsize max}}$ depends on $g/\Omega_{0}$ substantially.}
\label{Figure15}
\end{figure}

Figure~\ref{Figure13} shows variation of $t_{\mbox{\scriptsize max}}$,
when $p(t)$ becomes maximum,
with $T$ for $g/\Omega_{0}=1/4$.
Figure~\ref{Figure14} gives calculated $t_{\mbox{\scriptsize max}}$ versus $\gamma$ with putting $g/\Omega_{0}=1/4$.
Looking at Figs.~\ref{Figure13} and \ref{Figure14}, we notice $t_{\mbox{\scriptsize max}}$ move in a negative direction as $T$ and $\gamma$ become larger.
Figure~\ref{Figure15} shows $t_{\mbox{\scriptsize max}}$ as a function of $\gamma$ with putting $T=5.0\times 10^{-5}$~s.
Turning our eyes to Fig.~\ref{Figure15}, we perceive that $t_{\mbox{\scriptsize max}}$ moves in a negative direction as $g/\Omega_{0}$ decreases.

From Fig.~\ref{Figure15}, we can derive the following notion.
If we fix the parameters $\gamma$ and $T$ to specific values,
$t_{\mbox{\scriptsize max}}$ depends on $g/\Omega_{0}$ considerably.
Making $g/\Omega_{0}$ be smaller,
we observe that $t_{\mbox{\scriptsize max}}$ decreases drastically.
Thus, elevation of the pump intensity $\Omega_{0}$ of the classical light lets the generation of the single photon be earlier.
The trigger pulse causes the generation of the single photon.
However, the emission of the photon precedes the pump pulse.
This phenomenon seems interesting and strange.

\section{\label{section-discussion}Discussion}
In the current paper, we study the on-demand single-photon source implemented with the atom-cavity system by solving the master equation numerically.
As shown in Eq.~(\ref{half-time-T-0}),
the fluctuation of the duration of the emission has the upper bound, which is given by a linear function of $T$,
the characteristic time of the Gaussian trigger pulse.
Equation~(\ref{half-time-T-0}) also indicates that the fluctuation becomes smaller as $|\Omega_{0}/g|$ decreases.
However, even if $|\Omega_{0}/g|$ is equal to zero,
the upper bound of the fluctuation is given by $\delta t=\sqrt{2\ln 2}T$.
By contrast, the full width at half maximum of the Gaussian pump pulse is given by $2\tau=2\sqrt{\ln 2}T$.
If we let $g/\Omega_{0}=1/\sqrt{8}$,
the upper bound of the fluctuation becomes equal to the full width at half maximum of $\Omega(t)$.
Thus, adjusting the intensity of the pump pulse and the coupling constant of the Jaynes-Cummings interaction,
we can obtain single-photon emission which is narrower than the pump pulse.
This is one of merits that the scheme of Kuhn {\it et al}. has \cite{Kuhn1999}.

In Sect.~\ref{section-time-fluctuation},
we argue the trade-offs between the efficiency of generation of the single photon and the fluctuation of the duration of the emission.
These results restrict the performance of the single-photon gun in the laboratory.
Although the scheme of Kuhn {\it et al}. includes the excellent ideas, such as the use of the adiabatic process,
we have to know its limitations.
However, we do not  need to be too pessimistic because Fig.~\ref{Figure10} tells us
that the non-zero decay rate reduces the fluctuation of the duration of the emission.

In the present paper, we cannot find a function that approximates to $\delta t$ for non-zero $\gamma$.
We hope to solve this problem in the near future.

\appendix

\section{\label{section-numerical-calculation-master-equation}How to solve the master equation numerically}
In this section, we explain how to solve the master equation numerically.
First of all, for the sake of simplicity,
we introduce the following notation to describe the ket vectors:
\begin{eqnarray}
|0\rangle
&=&
|u,0\rangle, \nonumber \\
|1\rangle
&=&
|e,0\rangle, \nonumber \\
|2\rangle
&=&
|g,1\rangle, \nonumber \\
|3\rangle
&=&
|g,0\rangle.
\end{eqnarray}
We write down the density operator as
\begin{equation}
\rho(t)
=
\sum_{i=0}^{3}
\sum_{j=0}^{3}
C_{i,j}|i\rangle\langle j|,
\label{density-operator-definition}
\end{equation}
\begin{equation}
C_{j,i}^{*}(t)=C_{i,j}(t),
\label{coefficient-definition-1}
\end{equation}
\begin{equation}
\sum_{i=0}^{3}C_{i,i}=1.
\label{coefficient-definition-2}
\end{equation}
Then, we obtain the following first-order ordinary differential equation:
\begin{equation}
\frac{d}{dt}C_{i,j}(t)=\langle i|\frac{d}{dt}\rho(t)|j\rangle.
\label{differential-equation-0}
\end{equation}

On substitution from Eqs.~(\ref{master-equation-0}), (\ref{density-operator-definition}), (\ref{coefficient-definition-1}), and (\ref{coefficient-definition-2}),
Eq.~(\ref{differential-equation-0}) becomes
\begin{eqnarray}
\dot{C}_{0,0}
&=&
(i/2)\Omega(t)C_{0,1}-(i/2)\Omega(t)C_{1,0}, \nonumber \\
\dot{C}_{0,1}
&=&
(i/2)\Omega(t)C_{0,0}+i\Delta C_{0,1}+ig C_{0,2}-(i/2)\Omega(t)C_{1,1}, \nonumber \\
\dot{C}_{0,2}
&=&
ig C_{0,1}-(1/2)\gamma C_{0,2}-(i/2)\Omega(t)C_{1,2}, \nonumber \\
\dot{C}_{0,3}
&=&
-(i/2)\Omega(t)C_{1,3}, \nonumber \\
\dot{C}_{1,1}
&=&
-(i/2)\Omega(t)C_{0,1}+(i/2)\Omega(t)C_{1,0}+ig C_{1,2}-ig C_{2,1}, \nonumber \\
\dot{C}_{1,2}
&=&
-(i/2)\Omega(t)C_{0,2}+ig C_{1,1}+[-i\Delta-(1/2)\gamma] C_{1,2}-ig C_{2,2}, \nonumber \\
\dot{C}_{1,3}
&=&
-(i/2)\Omega(t)C_{0,3}-i\Delta C_{1,3}-ig C_{2,3}, \nonumber \\
\dot{C}_{2,2}
&=&
-ig C_{1,2}+ig C_{2,1}-\gamma C_{2,2}, \nonumber \\
\dot{C}_{2,3}
&=&
-ig C_{1,3}-(1/2)\gamma C_{2,3}, \nonumber \\
\dot{C}_{3,3}
&=&
\gamma C_{2,2}.
\end{eqnarray}

Here, we define fifteen real variables as follows:
\begin{eqnarray}
V_{1}&=&C_{0,0},
\quad
V_{2}=\mbox{Re}[C_{0,1}],
\quad
V_{3}=\mbox{Im}[C_{0,1}], \nonumber \\
V_{4}&=&\mbox{Re}[C_{0,2}],
\quad
V_{5}=\mbox{Im}[C_{0,2}],
\quad
V_{6}=\mbox{Re}[C_{0,3}], \nonumber \\
V_{7}&=&\mbox{Im}[C_{0,3}],
\quad
V_{8}=C_{1,1},
\quad
V_{9}=\mbox{Re}[C_{1,2}], \nonumber \\
V_{10}&=&\mbox{Im}[C_{1,2}],
\quad
V_{11}=\mbox{Re}[C_{1,3}],
\quad
V_{12}=\mbox{Im}[C_{1,3}], \nonumber \\
V_{13}&=&C_{2,2},
\quad
V_{14}=\mbox{Re}[C_{2,3}],
\quad
V_{15}=\mbox{Im}[C_{2,3}].
\end{eqnarray}
We let $\mbox{\boldmath $V$}$ denote a column vector with elements $\{V_{1}, ..., V_{15}\}$.
Then, we obtain the following system of differential equations:
\begin{equation}
\dot{\mbox{\boldmath $V$}}=\mbox{\boldmath $L$}\mbox{\boldmath $V$},
\label{differential-equation-1}
\end{equation}
where $\mbox{\boldmath $L$}$ is a $15\times 15$ matrix.
The elements of $\mbox{\boldmath $L$}$ are given by
\begin{eqnarray}
L_{1,3}&=&-\Omega(t), \nonumber \\
L_{2,3}&=&L_{10,9}=L_{12,11}=-\Delta, \nonumber \\
L_{2,5}&=&L_{4,3}=L_{10,13}=L_{12,14}=L_{15,11}=-g, \nonumber \\
L_{3,1}&=&L_{4,10}=L_{6,12}=L_{9,5}=L_{11,7}=(1/2)\Omega(t), \nonumber \\
L_{3,2}&=&L_{9,10}=L_{11,12}=\Delta, \nonumber \\
L_{3,4}&=&L_{5,2}=L_{10,8}=L_{11,15}=L_{14,12}=g, \nonumber \\
L_{3,8}&=&L_{5,9}=L_{7,11}=L_{10,4}=L_{12,6}=-(1/2)\Omega(t), \nonumber \\
L_{4,4}&=&L_{5,5}=L_{9,9}=L_{10,10}=L_{14,14}=L_{15,15}=-(1/2)\gamma, \nonumber \\
L_{8,3}&=&\Omega(t), \nonumber \\
L_{8,10}&=&-2g, \nonumber \\
L_{13,10}&=&2g, \nonumber \\
L_{13,13}&=&-\gamma, \nonumber \\
L_{i,j}&=&0\quad\mbox{for others}.
\end{eqnarray}

In the present paper, we numerically solve the system of differential equations (\ref{differential-equation-1})
by the Runge-Kutta method as follows:
\begin{eqnarray}
f_{i}(\mbox{\boldmath $L$}(t),\mbox{\boldmath $V$}(t))
&=&
\sum_{j=1}^{15}L_{ij}(t)V_{j}(t), \nonumber \\
(\mbox{\boldmath $k$}_{1})_{i}
&=&
f_{i}(\mbox{\boldmath $L$}(t),\mbox{\boldmath $V$}(t)), \nonumber \\
(\mbox{\boldmath $k$}_{2})_{i}
&=&
f_{i}(\mbox{\boldmath $L$}(t+\frac{\Delta t}{2}),\mbox{\boldmath $V$}(t+\frac{\Delta t}{2})+\frac{\Delta t}{2}\mbox{\boldmath $k$}_{1}), \nonumber \\
(\mbox{\boldmath $k$}_{3})_{i}
&=&
f_{i}(\mbox{\boldmath $L$}(t+\frac{\Delta t}{2}),\mbox{\boldmath $V$}(t+\frac{\Delta t}{2})+\frac{\Delta t}{2}\mbox{\boldmath $k$}_{2}), \nonumber \\
(\mbox{\boldmath $k$}_{4})_{i}
&=&
f_{i}(\mbox{\boldmath $L$}(t+\Delta t),\mbox{\boldmath $V$}(t+\Delta t)+\Delta t\mbox{\boldmath $k$}_{3}), \nonumber \\
V_{i}(t+\Delta t)
&=&
\frac{\Delta t}{6}
[
(\mbox{\boldmath $k$}_{1})_{i}
+
2(\mbox{\boldmath $k$}_{2})_{i}
+
2(\mbox{\boldmath $k$}_{3})_{i}
+
(\mbox{\boldmath $k$}_{4})_{i}
].
\label{Runge-Kutta-0}
\end{eqnarray}
To carry out numerical calculations actually, we let $\Delta t=T\times 2.0\times 10^{-6}$,
where $T$ is the characteristic time of the Gaussian trigger pulse defined in Eq.~(\ref{trigger-pulse-definition-0}).

\end{document}